\documentclass[12pt]{article}
\pdfoutput=1
\usepackage[utf8]{inputenc}
\usepackage{amsmath}
\usepackage{amssymb}
\usepackage{units}
\usepackage{graphicx}
\usepackage{slashed}
\usepackage{array}
\usepackage{comment}
\usepackage[leftcaption,raggedright]{sidecap}
\usepackage{caption}
\usepackage{wrapfig}
\usepackage{subcaption}
\usepackage[hidelinks]{hyperref}
\usepackage{cleveref}

\newcommand\pubnumber{NuPhys2026-Elizabeth-Long}
\newcommand\pubdate{\today}

\def\napoli{Charles University, Prague}

\newcommand\support{\footnote{
  A.~Akmete, R.~Aliberti, F.~Ambrosino, R.~Ammendola, B.~Angelucci, A.~Antonelli, G.~Anzivino, R.~Arcidiacono, M.~U.~Ashraf,
T.~Bache
% ...
}}

\def\Title#1{\begin{center} {\Large #1 } \end{center}}
\def\Author#1{\begin{center}{ \sc #1} \end{center}}
\def\Address#1{\begin{center}{ \it #1} \end{center}}

\newcommand\pubblock{\rightline{\begin{tabular}{l} \pubnumber\\
         \pubdate  \end{tabular}}}
\newenvironment{Abstract}{\begin{quotation}  }{\end{quotation}}
\newenvironment{Presented}{\begin{quotation} \begin{center} 
             PRESENTED AT\end{center}\bigskip 
      \begin{center}\begin{large}}{\end{large}\end{center} \end{quotation}}

\def\beq{\begin{equation}}
\def\eeq#1{\label{#1}\end{equation}}
\def\eeqn{\end{equation}}

\def\beqa{\begin{eqnarray}}
\def\eeqa#1{\label{#1}\end{eqnarray}}
\def\eeqan{\end{eqnarray}}

\let\bar=\overbar

\def\Dslash{\not{\hbox{\kern-4pt $D$}}}
\def\dslash{\not{\hbox{\kern-2pt $\del$}}}

\def\msb{{\bar{\ssstyle M \kern -1pt S}}}

\begin{document}
\begin{titlepage}
\pubblock

\vfill
\Title{New physics searches at NA62}
\vfill
\Author{Elizabeth Long for the NA62 collaboration\support}
\Address{\napoli}
\vfill
\begin{Abstract}
The NA62 experiment at CERN has collected a large sample of $K^+$ and $\pi^+$ decays in flight during Run 1 in 2016--2018 and the ongoing Run 2 which started in 2021. Searches for the decays $K^+\rightarrow\pi^+X$ and $\pi^+\rightarrow e^+N$ are presented using NA62 data collected in 2016--2022 and 2017--2024, respectively. Results are interpreted to constrain a range of new physics scenarios covering all four portal model scenarios. Upper limits on the $K^+\rightarrow\pi^+X$ branching ratio are established at the $10^{-11}$ level, providing constraints on dark photon, scalar and ALP couplings. From the search for heavy neutral lepton production in $\pi^+\rightarrow e^+N$ decays of beam pions, upper limits of the extended neutrino mixing matrix element $|U_{e4}|^2$ are established at the $10^{-8}$ level over the heavy neutral lepton mass range 95--126~MeV/$c^2$. 

\end{Abstract}
\vfill
\begin{Presented}
NuPhys2026, Prospects in Neutrino Physics\\
King's College, London, UK,\\ January 7--9, 2026
\end{Presented}
\vfill
\end{titlepage}
\def\thefootnote{\fnsymbol{footnote}}
\setcounter{footnote}{0}

\section{Introduction}
NA62 was designed to measure the branching ratio of the flavour-changing neutral current decay $K^+\rightarrow\pi^+\nu\bar{\nu}$. Using the SPS beam of 400 GeV protons, a secondary beam of 75 GeV particles is produced at a secondary target. This secondary beam is composed of 70\% $\pi^+$, 23\% p and 6\% $K^+$. In the 8 years of data taking so far, NA62 has accumulated the largest sample of $K^+$ decays ever collected, allowing for both high precision measurements of the Standard Model $K^+\rightarrow\pi^+\nu\bar{\nu}$ process, and searches for new physics such as hidden sectors and heavy neutral leptons.

\section{\texorpdfstring{$K^+\rightarrow\pi^+\nu\bar{\nu}$ decay}{Kpinn decay}}
The $K^+\rightarrow\pi^+\nu\bar{\nu}$ decay is highly suppressed due to the GIM mechanism and CKM suppression, giving it a Standard Model branching ratio $\mathcal{B}(K^+\rightarrow\pi^+\nu\bar{\nu})\sim8\times10^{-11}$ \cite{Buras2015,Buras2022}. Combined with the relatively clean theoretical predictions of the branching ratio of the decay, the low branching ratio makes this decay an important candidate to search for new physics.

\begin{figure}[h]
    \centering
    \includegraphics[width=0.9\linewidth]{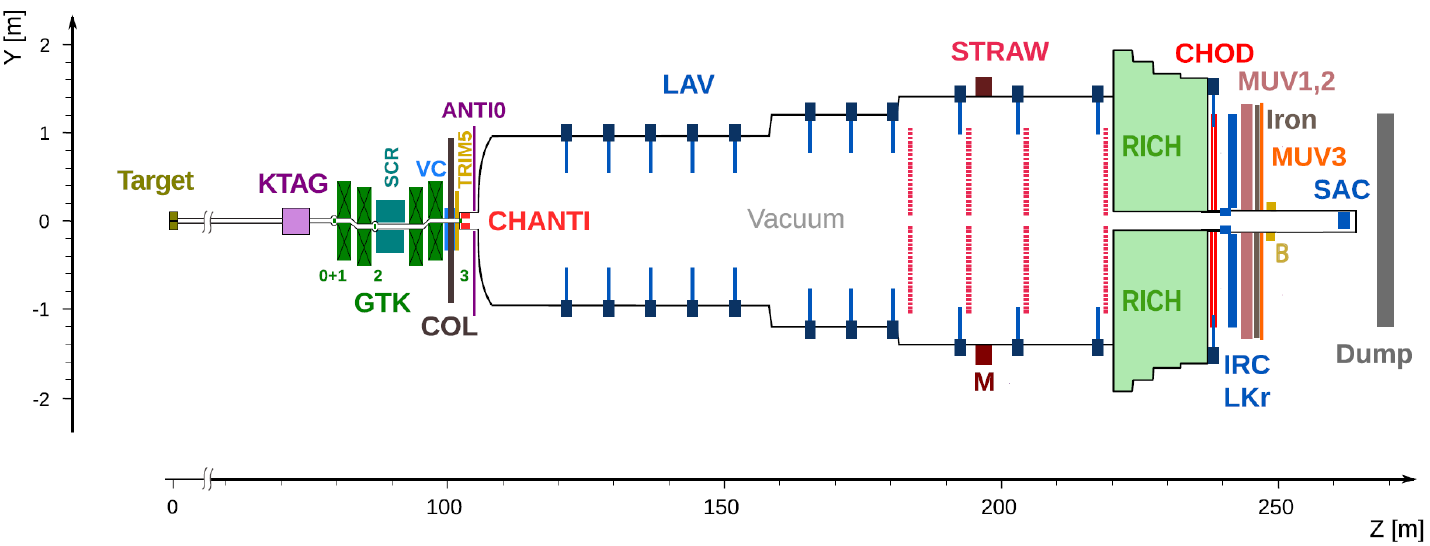}
    \caption{Schematic of NA62 experimental setup.}
    \label{fig:NA62Schematic}
\end{figure}

The experimental strategy relies on identifying a final state pion coming from a kaon which has decayed in the fiducial volume. The KTAG is used to tag the incoming kaon, and the GTK measures its momentum. The outgoing pion momentum is measured in the STRAW spectrometer, and the liquid krypton electromagnetic calorimeter (LKr) is used in conjunction with information from the muon vetos MUV1,2,3 and the RICH to identify the pion and reject background from muonic kaon decays. The RICH also provides timing information, and any additional activity is rejected using the large-angle veto (LAV).
A schematic of the experimental setup can be seen in \Cref{fig:NA62Schematic} and the full analysis strategy is discussed in \cite{NA62:2024pjp}.

\begin{figure}[h]
    \centering
    \includegraphics[height=0.35\textheight]{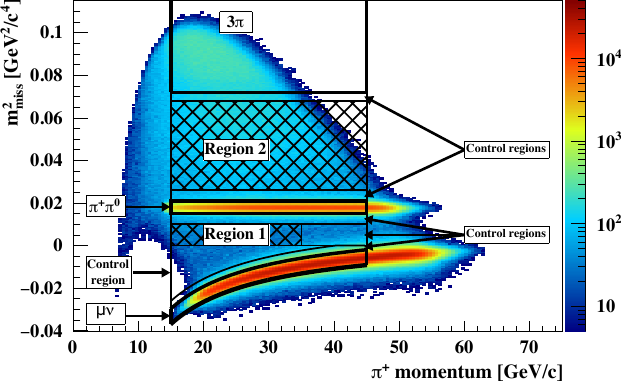}
    \caption{Reconstructed squared missing mass ($m^2_{miss}$) as a function of pion momentum ($\pi^+$ momentum) . Two signal regions (hatched areas) and control regions are shown. Also presented are the regions of $K\rightarrow3\pi$, $K\rightarrow\pi^+\pi^0$ and $K\rightarrow\mu\nu$ backgrounds.}
    \label{fig:mm2vsp}
\end{figure}

\newpage
\Cref{fig:mm2vsp} shows the kinematic distribution of kaon decays from a control data sample, before applying particle identification (PID) cuts and photon rejection. Choosing the signal regions to avoid the three main kaon decays ($K\rightarrow3\pi$, $K\rightarrow\pi^+\pi^0$ and $K\rightarrow\mu\nu$) results in background rejection of O($10^{-4}$). Additional PID cuts reduce background from the muonic decay by a further factor of O($10^8$), giving a single event sensitivity $\mathcal{B}_{\mathrm{SES}}=\left(N_{\mathrm{K}} A_{\pi \nu \bar{\nu}} \varepsilon_{\mathrm{trig}} \varepsilon_{\mathrm{RV}}\right)^{-1}=(8.48 \pm 0.29) \times 10^{-12}$, where: $A_{\pi \nu \bar{\nu}}=(7.62 \pm 0.22) \%$ is the signal selection acceptance, $\varepsilon_{\text {trig }}=(85.9 \pm 1.4) \%$ is the trigger efficiency ratio of the trigger lines used for the signal and normalisation samples and $\varepsilon_{\mathrm{RV}}=(63.2 \pm 0.6) \%$ is the random veto efficiency.

\begin{figure}[h]
        \begin{minipage}[t]{0.48\textwidth}
		\centering
        \includegraphics[height=0.29\textheight]{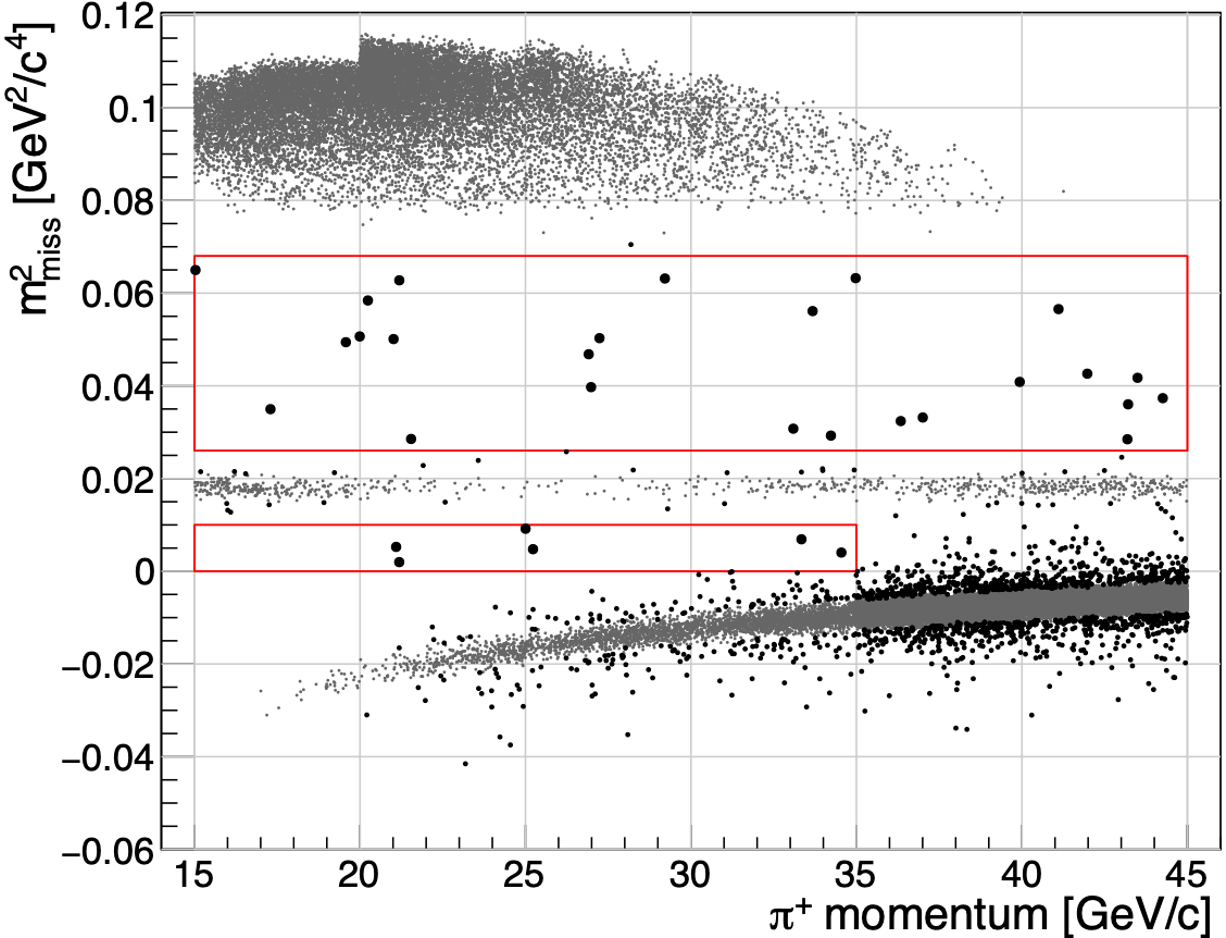}
        \caption{Squared missing mass ($m^2_{miss}$ as a function of pion momentum for events passing signal selection. The signal regions are highlighted in red.}
        \label{fig:PNNUnmasking}
	\end{minipage}%
        \hfill
	\begin{minipage}[t]{0.48\textwidth}
	    \centering
        \includegraphics[height=0.29\textheight]{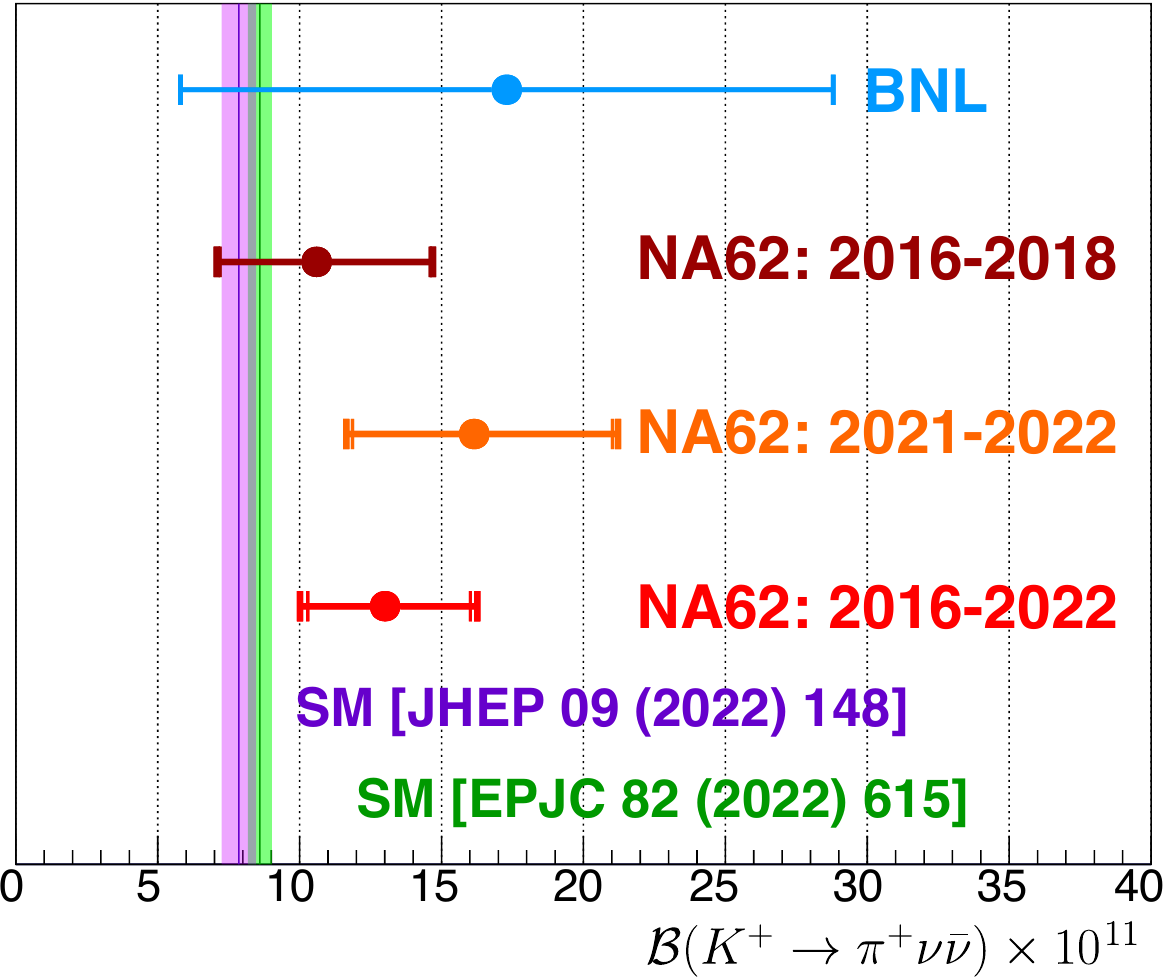}
        \caption{History of measurements and recent predictions of the $K^+\rightarrow\pi^+\nu\bar\nu$ branching fraction.}
        \label{fig:PNNincontext}
        \end{minipage}
\end{figure}

As seen \Cref{fig:PNNUnmasking}, in the full set of data taken between 2016-2022 after unmasking the signal regions, 51 signal events were found with an expected background of $18^{+3}_{-2}$, and the background-only hypothesis was rejected with $>5\sigma$ significance. The branching ratio was measured to be $\mathcal{B}(K^+\rightarrow\pi^+\nu\bar\nu) = \left(\left.\left.13.0^{+3.0}_{-2.7}\right|_{\text stat}{}^{+1.3}_{-1.3}\right|_{\text syst}\right)\times10^{-11}$. This result can be seen in the context of other experiments and the Standard Model in \Cref{fig:PNNincontext}. The result is in agreement with the Standard Model to within 1.7$\sigma$. Data taking for NA62 will continue in 2026, the total dataset is expected to be 3x the size of the sample collected between 2016-2022.

\section{Interpreting $K^+\rightarrow\pi^+\nu\bar\nu$ as $K^+\rightarrow\pi^+X$ search}
\begin{figure}[h]
    \begin{minipage}[t]{0.48\textwidth}
        \centering
        \includegraphics[width=\linewidth]{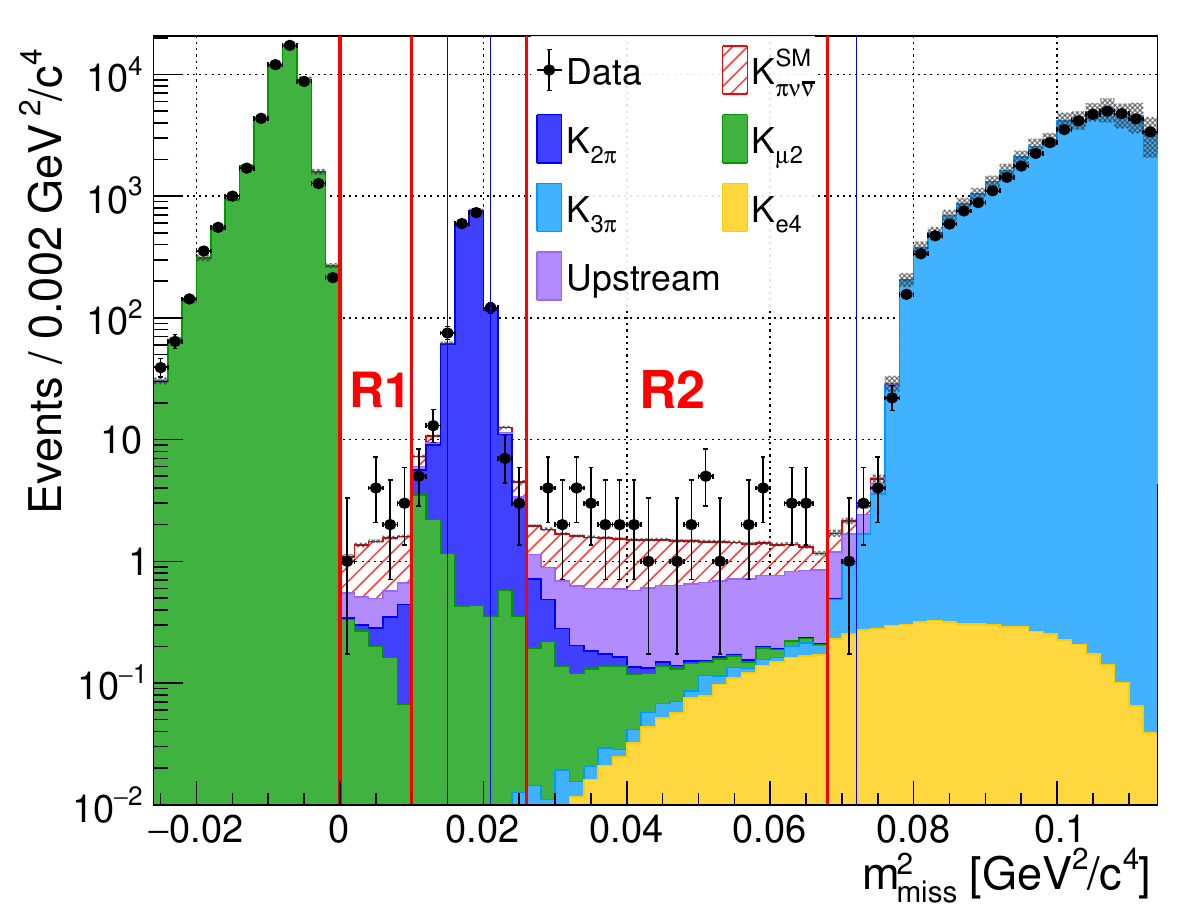}
        \caption{Missing mass spectrum of $K^+\rightarrow\pi^+\nu\bar\nu$ decay.}
        \label{fig:KpiX_2022_1a}
    \end{minipage}%
        \hfill
    \begin{minipage}[t]{0.48\textwidth}
        \centering
        \includegraphics[width=\linewidth]{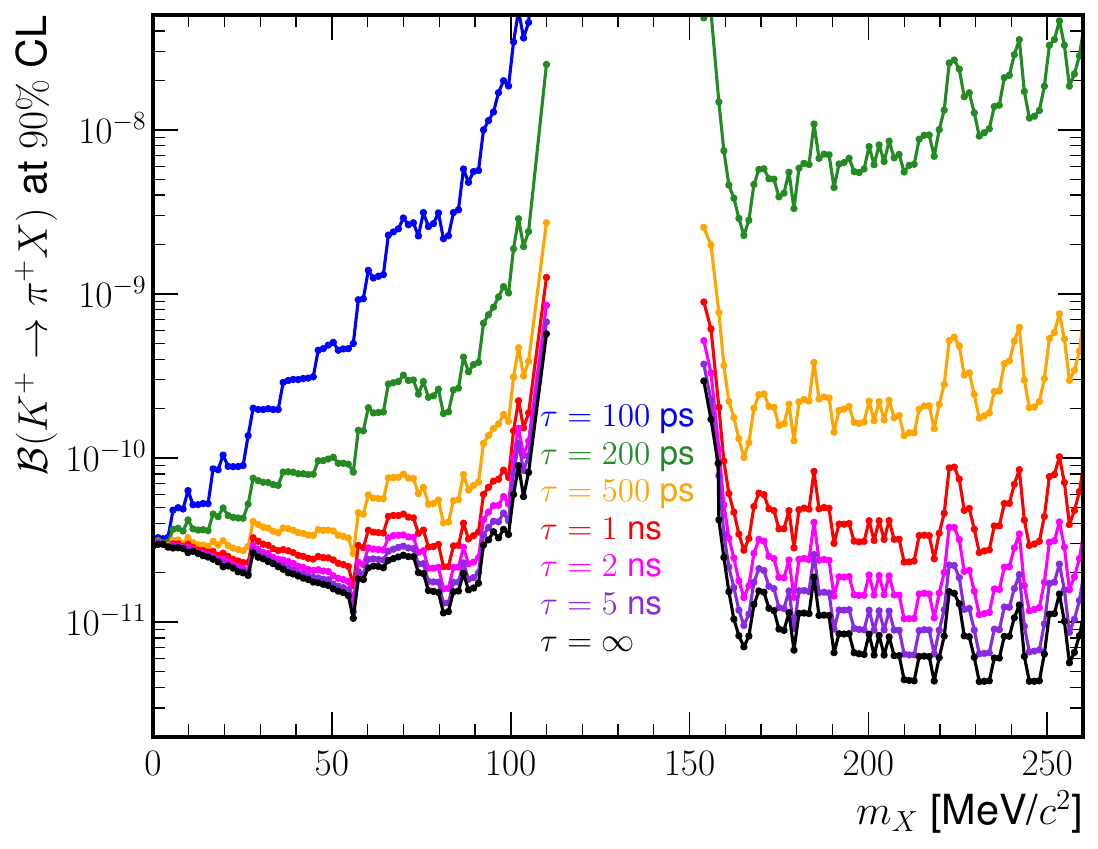}
        \caption{Model independent limits on branching fraction of $K^+\rightarrow\pi^+X$ for mass $m_X$ of particle with lifetime $\tau_X$.}
        \label{fig:KpiX_2022_2b}
    \end{minipage}
\end{figure}

\begin{figure}[h]
\centering
\begin{subfigure}[t]{0.48\textwidth}
    \centering
    \includegraphics[width=\linewidth]{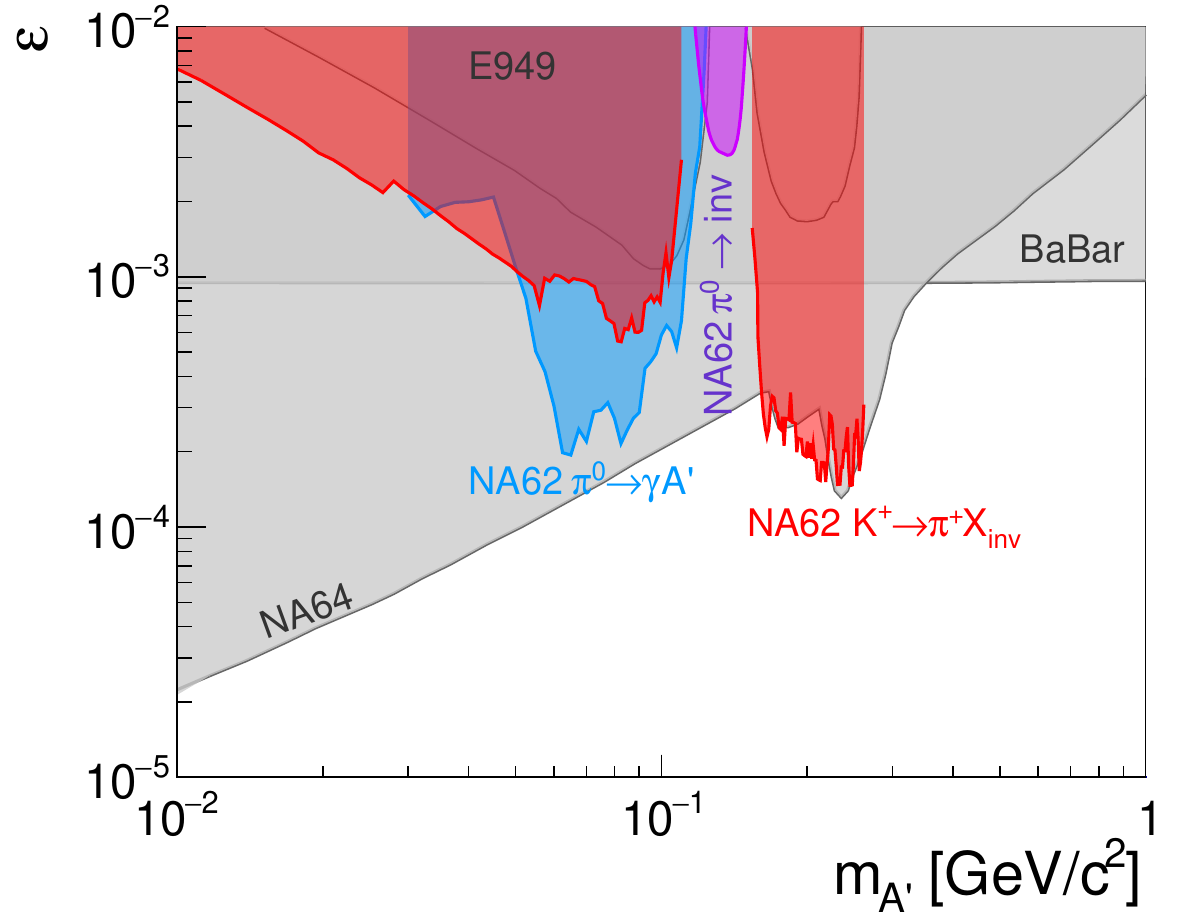}
    \caption{Vector interpretation}
\end{subfigure}
\hfill
\begin{subfigure}[t]{0.48\textwidth}
    \centering
    \includegraphics[width=\linewidth]{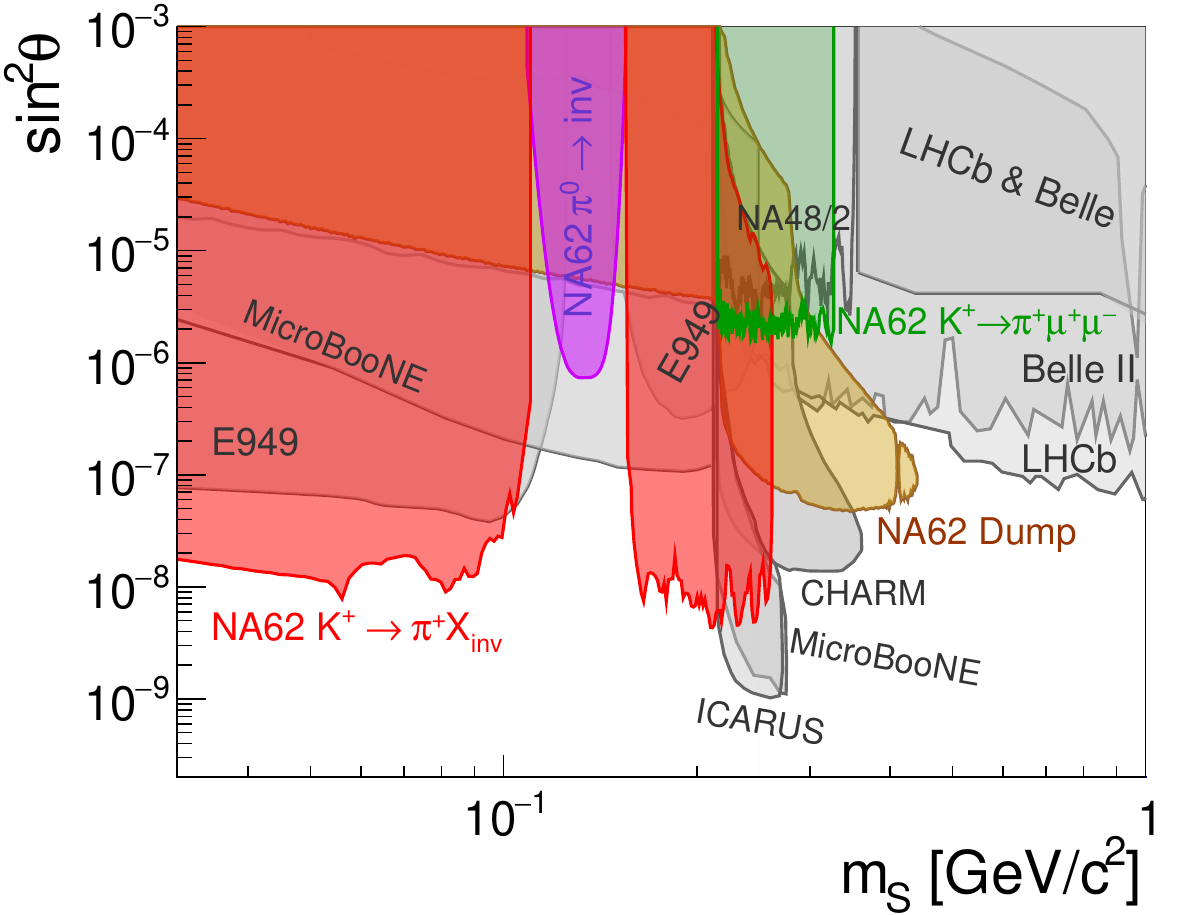}
    \caption{Scalar interpretation}
\end{subfigure}

\vspace{0.5cm}

\begin{subfigure}[t]{0.48\textwidth}
    \centering
    \includegraphics[width=\linewidth]{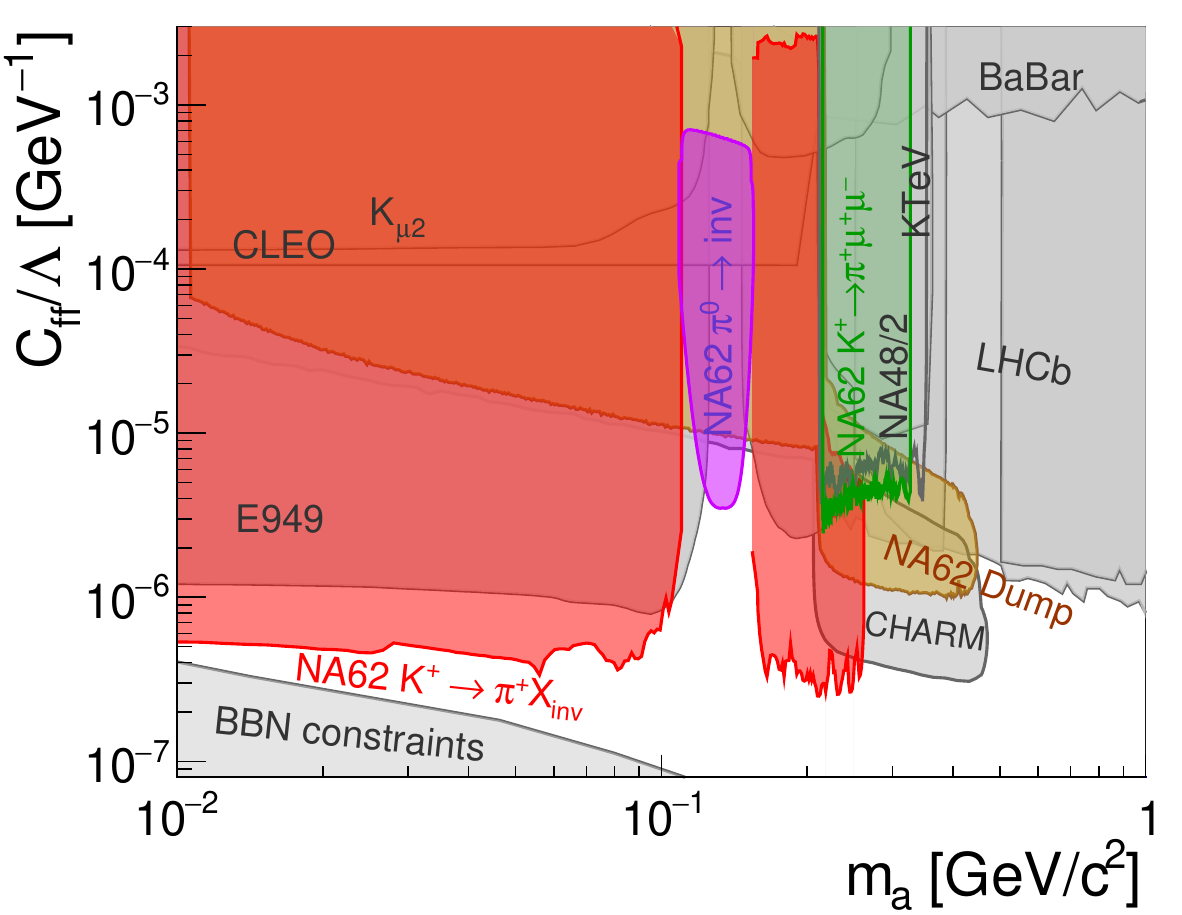}
    \caption{Interpretation as ALP with fermion coupling}
\end{subfigure}
\hfill
\begin{subfigure}[t]{0.48\textwidth}
    \centering
    \includegraphics[width=\linewidth]{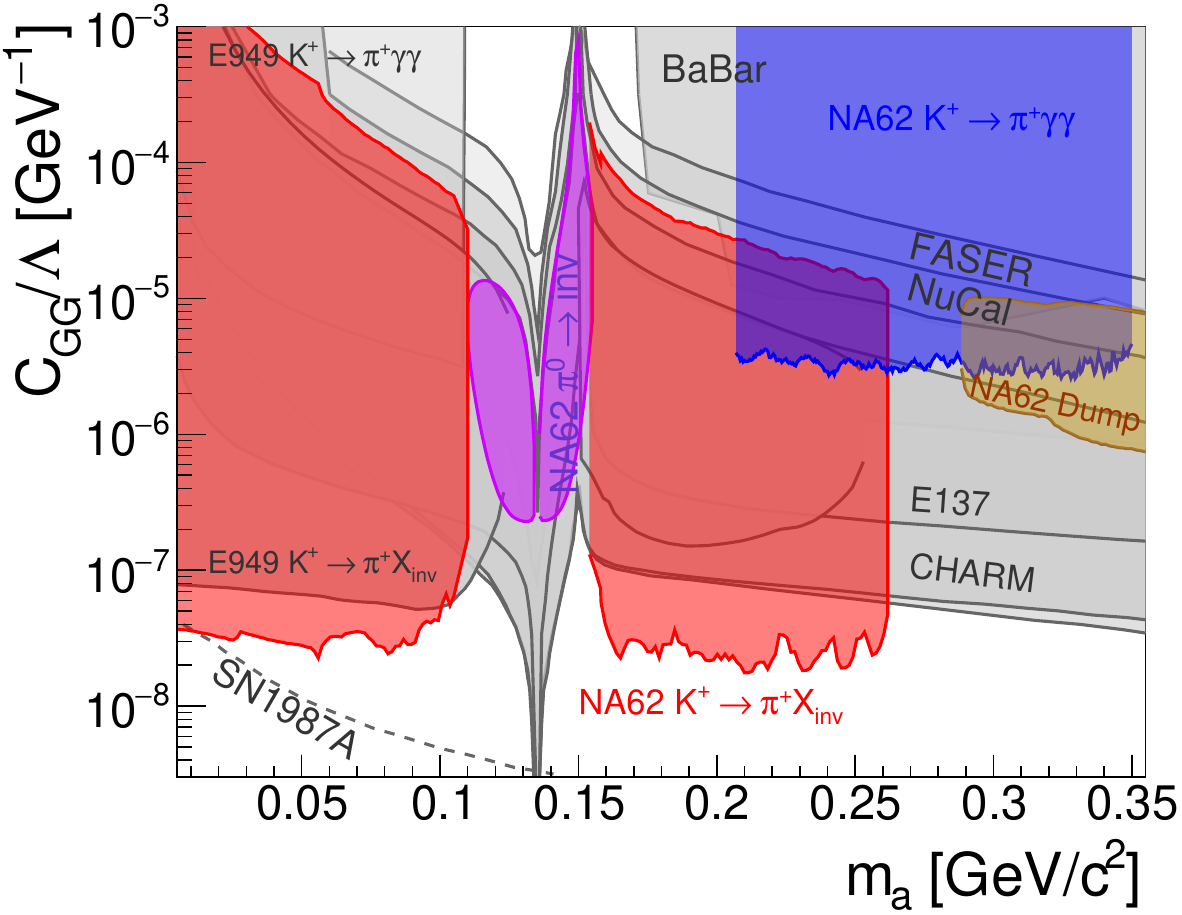}
    \caption{Interpretation as ALP with gluon coupling}
\end{subfigure}

\caption{Exclusion limits from NA62 searches for $K^+\rightarrow\pi^+X_{inv}$ assuming X is, respectively, a vector, a scalar, an ALP with fermion coupling, an ALP with gluon coupling.}
\label{fig:KPiXModelDependentExclusions}
\end{figure}

The $K^+\rightarrow\pi^+\nu\bar\nu$ has been reinterpreted as a search for a new BSM dark scalar or pseudo-scalar particle X in $K^+\rightarrow\pi^+X$ decays. The new particle X could either be stable, decay invisibly or decay into SM particles which are not detected by the experiment. In the last case, the signal of the decay would be identical to that of $K^+\rightarrow\pi^+\nu\bar\nu$, however the missing mass spectrum, shown in \Cref{fig:KpiX_2022_1a}, would contain a peak corresponding to the mass of the new particle. Performing a peak search on the missing mass spectrum, model-independent constraints are set for the case that X decays into visible particles with lifetime $\tau_X$, as shown in \Cref{fig:KpiX_2022_2b}. These limits are then reinterpreted as exclusion limits for the case that X is a vector, scalar, ALP with fermion coupling and ALP with gluon coupling respectively, as shown in \Cref{fig:KPiXModelDependentExclusions} \cite{NA62:2025upx}.

\newpage
\section{Heavy neutral lepton search $\pi^+\rightarrow e^+ N$}
The NA62 collaboration performed a search for heavy neutral leptons (HNLs) in electronic pion decays, where the signal would be identical to the Standard Model process $\pi^+\rightarrow e^+ \nu_e$.

The analysis used the same trigger line as for the $K^+\rightarrow\pi^+\nu\bar{\nu}$ analysis. The signal selection required a single positron in the final state and assumes that for HNLs with a lifetime $>$ 50 ns, produced in $\pi^+\rightarrow e^+ N$ decays with a Lorentz boost of $\mathcal{O}(500)$, any decays of the HNL into SM particles can be neglected.

\begin{figure}[h]
    \begin{minipage}[t]{0.48\textwidth}
		\centering
        \includegraphics[width=\linewidth]{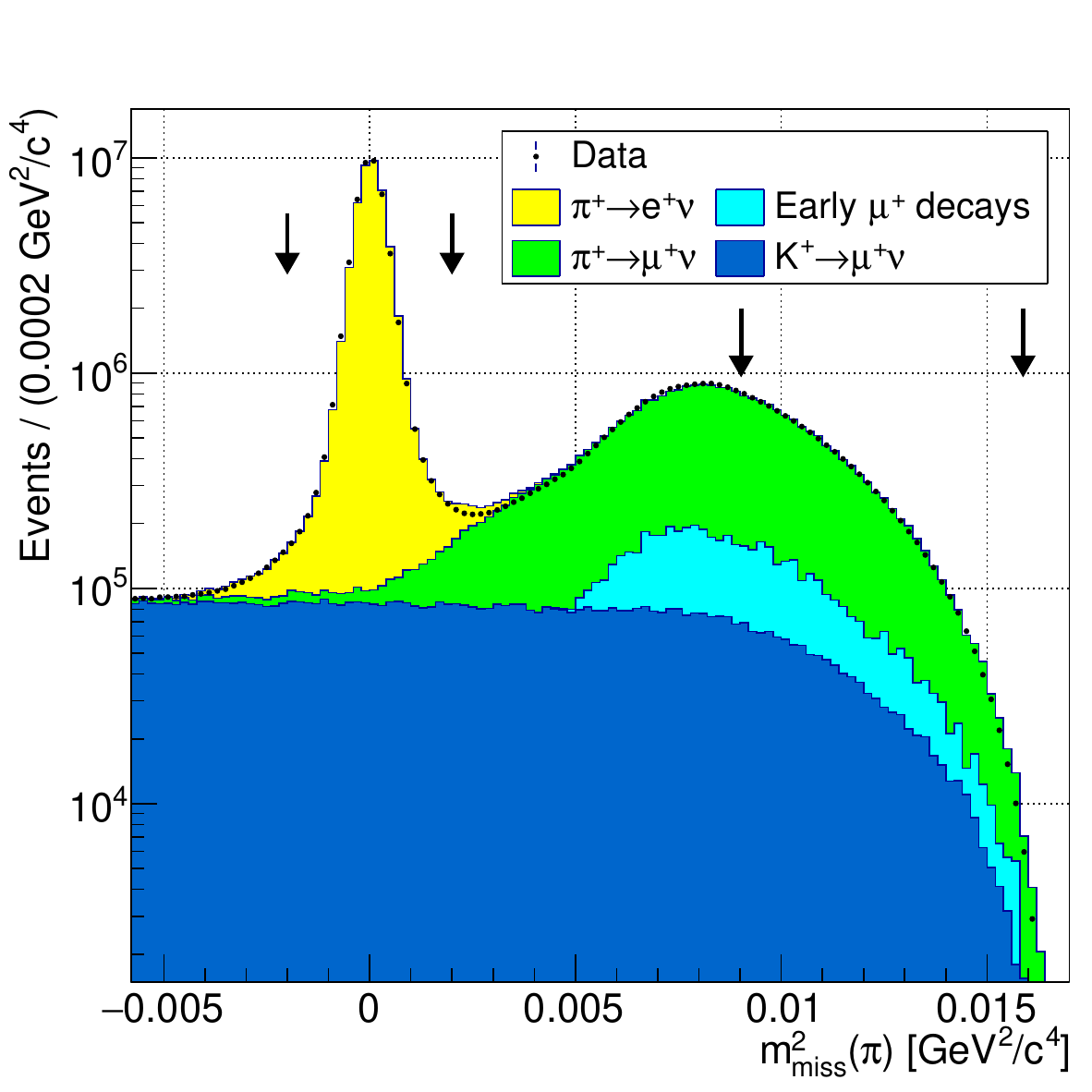}
        \caption{Squared missing mass distribution of electronic pion decays.}
        \label{fig:Fig2.1-Pi}
	\end{minipage}%
        \hfill
	\begin{minipage}[t]{0.48\textwidth}
	    \centering
        \includegraphics[width=\linewidth]{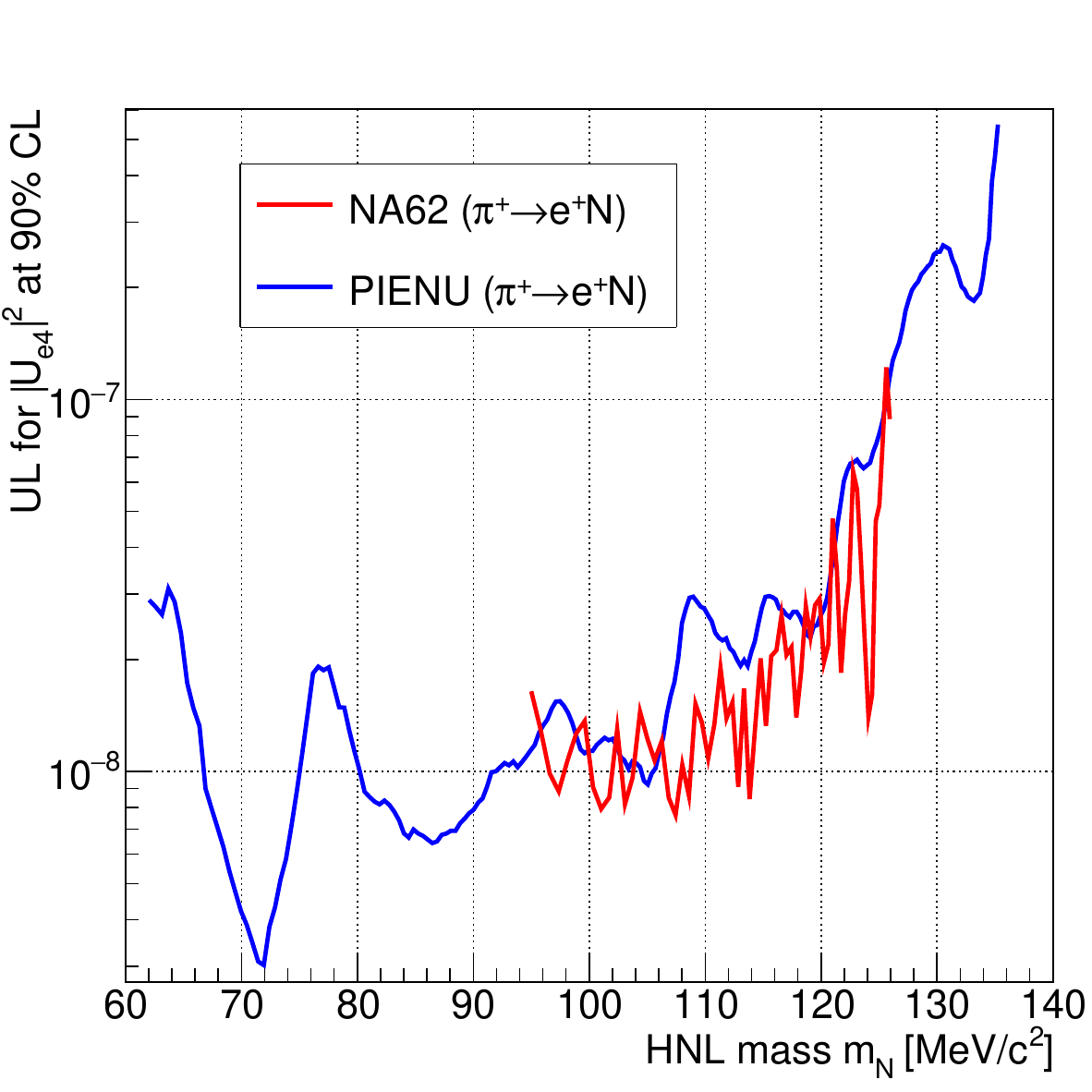}
        \caption{Upper limit on $|U_{e4}|^2$ mixing parameter as a function of HNL mass at 90\% confidence level}
        \label{fig:Fig5.1-ue4}
    \end{minipage}
\end{figure}

The search was carried out on data taken between 2017-2024 by doing a peak search in the missing mass spectrum $m^2_{miss}(\pi)$ shown in \Cref{fig:Fig2.1-Pi}, calculated as $(P_\pi-P_e)^2$, where $P_\pi$ and $P_e$ are the 4-momenta of the beam pion and decay electron respectively.  Here, the pair of arrows on the left, around the yellow peak, show the region used for the normalisation to $\pi^+\rightarrow e^+ \nu_e$ while the pair of arrows on the right show the $\pi^+\rightarrow e^+ N$ signal region \cite{NA62:2025csa}.

This search allowed upper limits to be set on the mixing parameter $|U_{e4}|$ between electrons and a heavy neutral lepton produced in this decay, as seen in \Cref{fig:Fig5.1-ue4}. At 90\% confidence level, limits were placed for $|U_{e4}|\sim10^{-8}$.

\newpage
\section{Conclusions and Outlook}
The NA62 collaboration has measured the branching ratio of $K^+\rightarrow\pi^+\nu\bar{\nu}$, showing agreement with the Standard Model to within 1.7$\sigma$. The $K^+\rightarrow\pi^+X$ search has allowed the collaboration to place competitive limits on all 4 dark sector portal models. The $\pi^+\rightarrow e^+ N$ search placed limits on $|U_{e4}|\sim10^{-8}$ at 90\% confidence level.

\newpage
\providecommand{\href}[2]{#2}\begingroup\raggedright\endgroup


\begin{thebibliography}{10}

\bibitem{Buras2015}
A.~J.~Buras, D.~Buttazzo, J.~Girrbach-Noe, and R.~Knegjens,
``K$^+ \to \pi^+ \nu \bar{\nu}$ and $K_L \to \pi^0 \nu \bar{\nu}$ in the Standard Model: status and perspectives,''
\textit{Journal of High Energy Physics}, vol.~2015, no.~11, p.~033, Nov.~2015.
\href{https://doi.org/10.1007/JHEP11(2015)033}{doi:10.1007/JHEP11(2015)033}.

\bibitem{Buras2022}
Buras, A.J., Venturini, E. The exclusive vision of rare K and B decays and of the quark mixing in the Standard Model. Eur. Phys. J. C 82, 615 (2022). \href{https://doi.org/10.1140/epjc/s10052-022-10583-8}{doi.org:10.1140/epjc/s10052-022-10583-8}

\bibitem{NA62:2024pjp}
E.~Cortina Gil \textit{et al.} [NA62],
JHEP \textbf{02} (2025), 191
doi:10.1007/JHEP02(2025)191
[arXiv:2412.12015 [hep-ex]].

\bibitem{NA62:2025upx}
E.~Cortina Gil \textit{et al.} [NA62],
JHEP \textbf{11} (2025), 143
doi:10.1007/JHEP11(2025)143
[arXiv:2507.17286 [hep-ex]].

\bibitem{NA62:2025csa}
B.~Bloch-Devaux \textit{et al.} [NA62],
Phys. Lett. B \textbf{872} (2026), 140119
doi:10.1016/j.physletb.2025.140119
[arXiv:2507.07345 [hep-ex]].

\end{thebibliography}
\end{document}